\documentclass[proof]{pasj00}

\usepackage{lscape}
\usepackage{threeparttable}

\begin{document}
\SetRunningHead{Kaneko et al.}{Properties of Molecular Gas in Galaxies in Early and Mid Stage of the Interaction: I}

\title{Properties of Molecular Gas in Galaxies in Early and Mid Stage of the Interaction: I. Distribution of Molecular Gas}

\author{Hiroyuki \textsc{Kaneko},\altaffilmark{1,2,3} Nario \textsc{Kuno},\altaffilmark{2,3} Daisuke \textsc{Iono},\altaffilmark{3}
Yoichi \textsc{Tamura},\altaffilmark{4} Tomoka \textsc{Tosaki},\altaffilmark{5} Koichiro \textsc{Nakanishi},\altaffilmark{2,6,7} and Tsuyoshi \textsc{Sawada}\altaffilmark{6,7} }

\altaffiltext{1}{Graduate School of Pure and Applied Sciences, University of Tsukuba, 1-1-1 Tennodai, Tsukuba, Ibaraki 305-8577}
\email{kaneko.hiroyuki.fw@u.tsukuba.ac.jp}
\altaffiltext{2}{Department of Astronomical Science, The Graduate University for Advanced Studies, 2-21-1 Osawa, Mitaka, Tokyo 181-8588}
\altaffiltext{3}{Nobeyama Radio Observatory, Minamimaki, Minamisaku, Nagano 384-1305}
\altaffiltext{4}{Institute of Astronomy, The University of Tokyo, 2-21-1, Osawa, Mitaka, Tokyo 181-0015}
\altaffiltext{5}{Department of Geoscience, Joetsu University of Education, Joetsu, Niigata 943-8512}
\altaffiltext{6}{National Astronomical Observatory of Japan, 2-21-1 Osawa, Mitaka, Tokyo 181-8588}
\altaffiltext{7}{Joint ALMA Observatory, Alonso de Cordova 3107, Vitacura, Santiago, Chile}


%

\KeyWords{galaxies: individual (Arp~84, VV~219, VV~254, the Antennae Galaxies) --- galaxies: interactions --- galaxies: ISM --- ISM: molecules} 

\maketitle

\begin{abstract}
We present the results of $^{12}$CO({\it J} = 1-0) mapping observations toward four interacting galaxies in early and mid stages of the interaction to understand the behavior of molecular gas in galaxy-galaxy interaction.
The observations were carried out using the 45-m telescope at Nobeyama Radio Observatory (NRO).
We compared our CO total flux to those previously obtained with single-dish observations and found that there are no discrepancy between them.
Applying a typical CO-H$_{2}$ conversion factor, all constituent galaxies have molecular gas mass more than 10$^{9}$ \MO.
Comparisons to H\emissiontype{I}, {\it Ks} and tracers of SF such as H$\alpha$, FUV, 8 $\mu$m and 24 $\mu$m revealed 
that the distribution of molecular gas in interacting galaxies in the early stage of the interaction differs from atomic gas, stars and star-forming regions.
These differences are not explained without the result of the interaction.
Central concentration of molecular gas of interacting galaxies in the early stage of the interaction is lower than that of isolated galaxies, which suggests molecular gas is distributed 
off-centre and/or extends in the beginning of the interaction.
\end{abstract}

\section{Introduction}
Galaxy-galaxy interactions including mergers and collisions play important roles in galaxy formation and evolution.
Close interaction with other galaxies not only changes the morphology and the kinematics of stars and gas in a galaxy \citep{TT72} but also enhances star-formation (SF) activity.
\citet{LT78} attribute large colour dispersion of interacting systems in colour-colour diagrams compared to isolated galaxies to bursts of SF in interacting galaxies.
Subsequent observational studies using tracers of SF such as H$\alpha$ \citep{Bushouse87,Kennicutt87}, far-infrared (FIR) \citep{Bushouse88}, mid-infrared (MIR) \citep{JW85,Lonsdale84} and radio continuum \citep{Stocke78,Sulentic76} all prove the enhancement of SF activity in interacting galaxies.
These global and statistical studies show that a galaxy interaction makes a great contribution to SF activity of galaxies.

SF occurs through a self-gravitational disruption of molecular gas clouds in interstellar space.
It is very important to know how interstellar matter, especially, molecular gas is affected by interactions of galaxies to understand the mechanism of active SF in interacting galaxies.
Since infrared (IR) luminous galaxies are expected to have strong CO emission, early CO observations toward extragalaxies were mainly focused on Luminous Infrared Galaxies (LIRGs) and Ultra Luminous Infrared Galaxies (ULIRGs) whose IR luminosity is larger than 10$^{11}$ and 10$^{12}$ \LO, respectively \citep{Soifer84}.
Although these observations revealed that LIRGs have a large amount of molecular gas ($>$ 10$^{10}$ \MO) and 
that there is a good correlation between CO luminosity and IR luminosity, 
the reasons for the enhanced SF activity have not been well understood mainly due to low spatial resolution (e.g., \cite{Young89,YS91}).

With the improvement of receiver sensitivity, it has been possible to observe normal galaxies in CO and to compare SF activity between LIRGs and normal galaxies.
The correlation between CO and IR luminosities was still found for normal galaxies and 
high star formation efficiency (SFE) was found in interacting galaxies compared to normal spiral galaxies \citep{Young86b,Sanders91,Combes94}. 
Considering these results, \citet{SS88} suggest that interacting galaxies form stars more effectively than normal galaxies.
Recent observational studies, however, found some facts that oppose this scenario.
Statistical studies of interacting galaxies show that molecular gas mass in interacting galaxies is higher than that in field galaxies \citep{BC93, Casasola04, Combes94, Zhu01}.
Moreover, \citet{Casasola04} also show that, contrary to the previous results, SFE is not higher in ULIRGs where star formation rate (SFR) is higher than field galaxies.
The discrepancies of SFE in previous papers occurred because past observations of molecular gas were pointed at only the centre of galaxies.
For such observations, it is difficult to get even total CO flux and a distribution of molecular gas in galaxies.

\citet{Zhu01} showed that the enhancement of SFR begins in the mid stage of the interaction and most active SF happens at the late stage of the interaction classified by their morphologies.
This fact suggests that the conditions of triggering starburst have been satisfied in earlier stages of the interaction.
The key to the onset of starburst is suggested by the numerical simulations by \citet{BH96}.
They revealed that gravitational torque makes interstellar gas infall into the central regions of progenitors during the interaction.
Since the density of interstellar gas increases at the central regions due to the gas infall, SF at the galactic centre takes place efficiently.
High resolution interferometric CO observations toward LIRGs clarified that molecular gas in LIRGs is concentrated in the central 1 kpc region \citep{Scoville91,Yun94}.
Since most LIRGs are regarded as being in the late stage of the interaction, measurements of the central concentration of molecular gas in earlier stage of the interaction are very important in order to explain when and how molecular gas falls into the nuclei.
Therefore, observations of interacting galaxies in earlier stages of the interaction are required.

To examine whether changes of molecular gas properties happen under the interaction, 
it is necessary to observe molecular gas which covers the entire region of interacting galaxies with an enough spatial resolution to resolve their main structure.
The 25-BEam Array Receiver System (BEARS) \citep{Sunada00} equipped on the 45-m telescope at NRO is one of the most suitable instruments for such observations. 
Thus we conducted $^{12}$CO({\it J} = 1-0) mapping observations of interacting galaxies using the 45-m telescope.

In this paper, we present the data from our $^{12}$CO({\it J} = 1-0) mapping observations toward four interacting galaxies in the early and the mid stages of the interaction.
We investigate the influence of the interaction through the comparison between a central concentration of molecular gas and that of stars.
The sample selection and the details of observations are described in section \ref{obs}.
The CO data and comparisons to previous observations are presented in section \ref{results}.
Central concentration of molecular gas and stars are discussed in section \ref{discussion}.
We will discuss molecular gas fraction and its relationship with star-forming activity in detail in forthcoming papers.

\section{Observations}
\label{obs}
\subsection{Target Selection}

Since our purpose is to study the properties of molecular gas in the early stages of the interaction, 
we selected interacting galaxies from the sample of \citet{Zhu01} based on the requirements as follows:\\\\
1. Stage of the interaction

To investigate the effects of the interaction on molecular gas in interacting galaxies and how SF is enhanced,
the target galaxy pairs are in the early and the mid stages of the interaction.
There are many classifications of interacting galaxies and most of them are determined by their optical morphologies. 
In this paper, we adopt the definition proposed by \citet{Zhu01}.
That is, interacting galaxies whose $R < 1.5D_{25}$ were adopted as in the early stage of the interaction, where {\it R} is the separation between the nuclei and $D_{25}$ is the diameter of the primary galaxy, 
using the brightness contour $B_{T}$ = 25 mag arcsec$^{-2}$, and showing small morphological disturbances.
Interacting galaxies which come into contact with companions in projected and which have severely disturbed morphology, i.e. colliding galaxies., are classified as mid stage of the interaction.\\\\
2. Intensity of CO emission

Typically, the CO emission from the galactic disc region is a few times weaker than that from the central region.
Therefore, we selected galaxy pairs in which CO emission was detected from the central region with 200 Jy km/s in previous single dish observations.\\\\
3. Size of galaxy pair

To resolve major structures, small interacting galaxies are not suitable for our study.
We choose galaxy pairs whose apparent size is as large as possible, for instance, 2.$\timeform{'}$5$\times$2.$\timeform{'}$5.\\

\begin{figure}
	\begin{center}
		\FigureFile(80mm,80mm){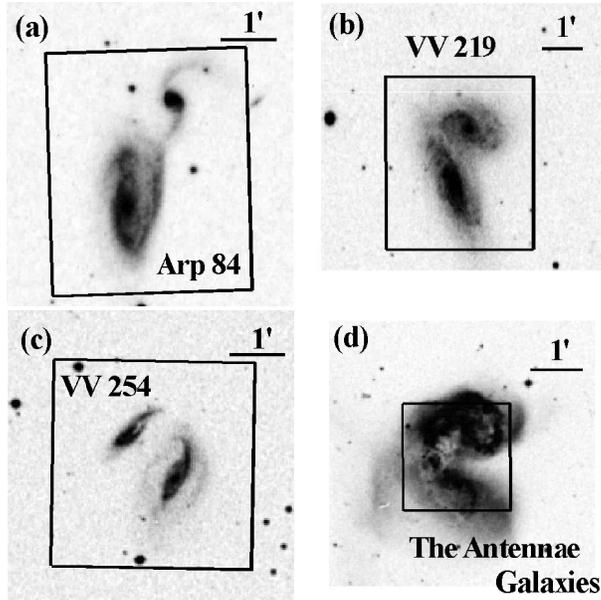}
		\caption{The DSS image of our sample galaxies. The rectangles indicate our mapping area (\timeform{3'.5} $\times$ \timeform{3'.5} for Arp~84, VV~219 and VV~254 and \timeform{2'.5}  $\times$ \timeform{2'.5} for the Antennae Galaxies).}
		\label{targets}
	\end{center}
\end{figure}

According to our requirements, we choose four interacting galaxies; Arp~84, VV~219, VV~254 and the Antennae Galaxies as the target galaxy pairs.
Basic information for each galaxy pair is described below.\\

\noindent 
Arp~84 (NGC~5394 and NGC~5395)

Arp~84 is an interacting galaxy pair in the early stage of the interaction with the distance of 53 Mpc and the systemic velocity of 3471 km s$^{-1}$.
Digital Sky Survey (DSS) image of Arp~84 is shown in Fig.\ref{targets}(a).
It consists of NGC~5394 (north-west) and NGC~5395 (south-east) with a projected separation of 38 kpc and colliding in a prograde-retrograde interaction.
\citet{Kaufman99} suggested that NGC~5394 is nearly face-on from the narrow H\emissiontype{I} line width and Arp~84 has undergone a first encounter from a numerical simulation.\\

\noindent
VV~219 (NGC~4567 and NGC~4568)

VV~219 galaxy pair is one of the nearest interacting galaxies located in the Virgo Cluster, for which we can resolve the main structures and the colliding area (see, Fig.\ref{targets}(b)).
The projected separation between two galaxies, NGC~4567 (north-west) and NGC~4568 (south-east),
 of 11 kpc and their weakly disturbed optical morphology (e.g., a tidal bridge, tails and warped disks) implies this galaxy pair is in a very early stage of interaction,
 probably the first encounter for this interacting system.
The distance of VV~219 derived from the systemic velocities, 2249 km s$^{-1}$, is 30 Mpc (using {\it H$_{0}$} = 73 km s$^{-1}$ Mpc$^{-1}$), 
but the distance of M 87 at the centre of the Virgo Cluster is given as 16 Mpc \citep{Mei07}.
This discrepancy may be due to the high peculiar velocity of VV~219 in the cluster potential. 
Since the fact of a projected distance between VV~219 and M 87 of 1 Mpc suggests that VV~219 is located in the inner region of the Virgo Cluster, we adopt 16 Mpc for the distance of VV~219.\\

\noindent
VV~254 (UGC~12914 and UGC~12915)

This pair is a head-on colliding galaxy pair in which the disks of the galaxies collide parallel as shown in Fig.\ref{targets}(c).
The distance to VV~254 is 62 Mpc which is the most distant interacting galaxies in our targets.
The galaxies located in the south-west and north-east are UGC~12914 and UGC~12915, respectively.  
The projected separation between two galaxies is 20 kpc and the systemic velocity is 4353 km s$^{-1}$.
The optical morphology of both UGC~12914 and UGC~12915 shows little disturbance.
VV~254 is known as Taffy I and have a unique feature, that is, synchrotron emission is seen in the bridge between the two galaxies \citep{Condon93}.
The direction of the linear polarization in the bridge region is aligned and perpendicular to the discs which suggests the trail of the collision.
Since interacting galaxies VV~769 (Taffy II) which show the same features collide in a head-on collision \citep{Condon02}, 
a head-on collision may induce aligned synchrotron emission with aligned polarization.
The mechanism of the synchrotron emission is still not clear.
\citet{Condon93} estimate the time elapsed since the collision for VV~254 attributing the synchrotron emission to the collision 
and conclude that the collision occurred $\sim$ 2 $\times$ $10^{7}$ years ago.
This means VV~254 experienced the head-on collision recently.\\

\noindent
The Antennae Galaxies (NGC~4038 and NGC~4039)

The Antennae Galaxies (Arp 244) are one of the representative prograde-prograde interacting galaxies and named for their elongated tidal tails.
A northern galaxy is NGC~4038 and a southern galaxy is NGC~4039.
The distance to the Antennae Galaxies is 21 Mpc \citep{Schweizer08}.
Since they are one of the nearest interacting galaxies and show active SF in their distribution of molecular gas region,
many observations have been carried out.
According to the definition by \citet{Zhu03}, the Antennae Galaxies are classified as interacting galaxies in mid stage of the interaction with a projected separation of 7 kpc.\\
 
In table \ref{target}, basic data of our target interacting galaxies are summarized. 
Following the pair data, basic data of constituent galaxies of the target are shown in table \ref{constituent}.

\begin{table*}
	\begin{center}
		\caption{Main parameters of galaxy pairs of our sample}
		\begin{tabular}{ccccc}
				\hline
				Name      & Velocity   & Distance &  Separation & Resolution \\
				          &       (km s$^{-1}$)          & (Mpc)                      &  (kpc)      & (kpc)       \\
 				(1)       &       (2)                    &        (3)                 &    (4)      &      (5)    \\
				\hline
				Arp~84    &                 3471         &       53                   &  38         & 5.0   \\
				VV~219    &                 2249         &       16                   &   11        & 1.5   \\
				VV~254    &        4353                  &       62                   &   20        & 5.8   \\
                Antennae Galaxies &         1705     &       21                   &   7         & 2.0   \\
				\hline
			\multicolumn{5}{@{}l@{}}{\hbox to 0pt{\parbox{180mm}{\footnotesize
         	   Column (1): Name of galaxies pairs. 
     		    \par\noindent
               Column (2): Heliocentric velocity in local standard of rest from NASA/IPAC Extragalactic Database(NED).
	     		\par\noindent
               Column (3): Distance of the galaxy pairs from NED.
                \par\noindent
               Column (4): Projected physical separation between two galaxies of a pair.
               	\par\noindent
               Column (5): Linear spatial resolution corresponding to the effective angular resolution, \timeform{19''.3},\\
	           at the distance of galaxy pairs.
			}\hss}}
			\label{target}
		\end{tabular}
	\end{center}
\end{table*}

\begin{table*}
	\begin{center}
		\caption{Main parameters of constituent galaxies of galaxy pairs}
		\tabcolsep2pt
		\small
	    \begin{tabular}{cccccccc}
				\hline
				Galaxy pair name & Constituent galaxies name & Right Ascension & Declination &  Morphology & Inclination & Orbit & $R_{\mathrm{K20}}$ \\
				&          &                        &                       &               &   (deg)     &       & (arcsec) \\
 				(1)       &       (2)              &        (3)            &    (4)        &     (5)     &  (6)  &   (7)  & (8)  \\
				\hline
				Arp~84 &NGC~5394  & \timeform{13h58m33.7s} & \timeform{+37D27'13"} & SB(s)b pec &      0      &   P   &    34    \\
				& NGC~5395  & \timeform{13h58m37.9s} & \timeform{+37D24'28"} & SA(s)b pec &     58      &   R   &    71    \\
				\hline 
				VV~219&NGC~4567  & \timeform{12h36m32.7s} & \timeform{+11D15'29"} & SA(rs)bc &     44      &   R   &    70    \\
				&NGC~4568  & \timeform{12h36m34.2s} & \timeform{+11D14'20"} & SA(rs)bc &     58      &   R   &    94    \\
				\hline 
				VV~254&UGC~12914 & \timeform{00h01m38.3s} & \timeform{+23D29'01"} & (R)S(r)cd pec &     61      &   H   &    55    \\
			    &UGC~12915 & \timeform{00h01m41.9s} & \timeform{+23D29'45"} &     S?        &     73      &   H   &    40    \\
				\hline
			    Antennae Galaxies & NGC~4038  & \timeform{12h01m53.0s} & \timeform{-18D52'03"} &  SB(s)m pec & --- &   P   &    73    \\
				&NGC~4039  & \timeform{12h01m53.5s} & \timeform{-18D53'10"} &  SA(s)m pec   &    ---      &   P   &    81    \\
				\hline
			\multicolumn{8}{@{}l@{}}{\hbox to 0pt{\parbox{180mm}{\footnotesize
                Column (1): Name of the constituent galaxy
			\par\noindent
                Column (2): Right ascension from NED.
			\par\noindent
                Column (3): Declination from NED
			\par\noindent
                Column (4): Morphological type from NED.
            \par\noindent
                Column (5): Inclination angle. Reference: \citet{Kaufman99} for Arp~84, this work for VV~219 and \\
				\citet{Giovanelli86} for VV~254. Since the Antennae Galaxies have complex morphologies, the inclination is not fixed and is not given.
            \par\noindent
                Column (6): Orbital type judging from their morphology; P, R, H represent a prograde, retrograde and head-on collision, \\
				respectively.
            \par\noindent
                Column (7): The fitted radius at 20 mag arcsec$^{-2}$ in the {\it Ks}-band.
			}\hss}}
			\label{constituent}
		\end{tabular}
	\end{center}
\end{table*}

\subsection{$^{12}$CO({\it J} = 1-0) Observations}
\label{COobservation}
$^{12}$CO observations were made toward the target interacting galaxies with the 45-m telescope at Nobeyama Radio Observatory (NRO)\footnote{NRO 45-m radio telescope is operated by Nobeyama Radio Observatory, a branch of National Astronomical Observatory of Japan.}.
We used BEARS which has twenty-five beams with a beam separation of \timeform{41''.1} \citep{Sunada00} as the receiver front-ends.
The typical system noise temperature during the observations was $T_{sys} \sim$ 250-650 K (in  double sideband mode) depending on mainly the weather.
The full width at half-power (FWHP) was \timeform{16''} at the rest frequency of $^{12}$CO({\it J} = 1-0)(115.271204 GHz).

We employed the On-The-Fly (OTF) mapping technique \citep{Sawada08}
to map the area of \timeform{3'.5} $\times$ \timeform{3'.5} which covers the whole interacting system
except for the Antennae Galaxies for which the central \timeform{2'.5} $\times$ \timeform{2'.5} region was observed.
The mapped regions for each galaxy pair are indicated in Fig.\ref{targets} with rectangles.
Throughout the observations, the data were sampled every 0.1 s.
The OFF points were taken at \timeform{7'} from the centre of the interacting system.

Twenty-five digital autocorrelators (AC45) \citep{Sorai00} were used as the back-ends.
The total band width is 512 MHz (1024 channels) which corresponds to velocity coverages of 1332 km s$^{-1}$ at 115 GHz.
The frequency resolution is 500 kHz which corresponds to a velocity resolution of 1.3 km s$^{-1}$ at 115 GHz. 

The telescope pointing accuracy was checked every hour by observing SiO maser lines 
from nearby evolved stars or 43 GHz continuum from bright QSOs with a SIS receiver (S40) or a HEMT receiver (H40).
The antenna temperature, $T_{A}^{*}$, was obtained by the chopper-wheel method correcting for atmospheric and antenna ohmic losses.
The main-beam efficiency of the telescope at 115 GHz was $\eta_{{\rm MB}}$ = 0.39 $\pm$ 0.03, 0.34 $\pm$ 0.03 and 0.31 $\pm$ 0.03 for 2004-2006, 2006-2007 and 2008-2009 seasons, respectively. 
The intensity given in this paper is the main-beam brightness temperature, which is defined by $T_{{\rm MB}} \equiv T_{{\rm A}}^{*}/\eta_{{\rm MB}}$.

\subsection{Data Reduction}
The data were reduced with the software packages NOSTAR
 (Nobeyama On-The-Fly Software Tools for Analysis Reduction) \citep{Sawada08}.
We removed the data whose pointing error was larger than \timeform{5''} to avoid the systematic intensity loss.
Since BEARS is a double sideband (DSB) receiver, 
we calibrated the sideband ratios of each beam of BEARS adopting the scaling factors 
derived from the observations of a bright standard source using both BEARS and the single sideband (SSB) receiver S100.
The calibration error was less than 5 \% for each array.
After the calibration, we made baseline fitting to remove linear baselines.
The data were convolved with a Gaussian-tapered Bessel function
\begin{equation}
	\frac{J_{1}(r/a)}{r/a}\times exp\left[ -\left( \frac{r}{b}\right)^{2}\right] \qquad a = 1.55/\pi, \ b = 2.52
\end{equation}
to create a cube data with \timeform{7''.5} grid spacing and a velocity resolution of 20 km s$^{-1}$.
In the equation, $J_{1}$ is the Bessel function of the first kind of order 1 and $r$ the distance between data point and grid point in pixel.
The effective angular resolution of the data is \timeform{19''.3}.
Finally, we fitted and subtracted linear baselines for each spectra again.
It is very important to get good baseline for observations of interacting galaxies because of their completely unpredicted distribution of molecular gas and wide velocity width of their CO line.

\section{Results}
\label{results}
\subsection{Spectrum maps}
\begin{figure*}
	\begin{center}
		\FigureFile(160mm,150mm){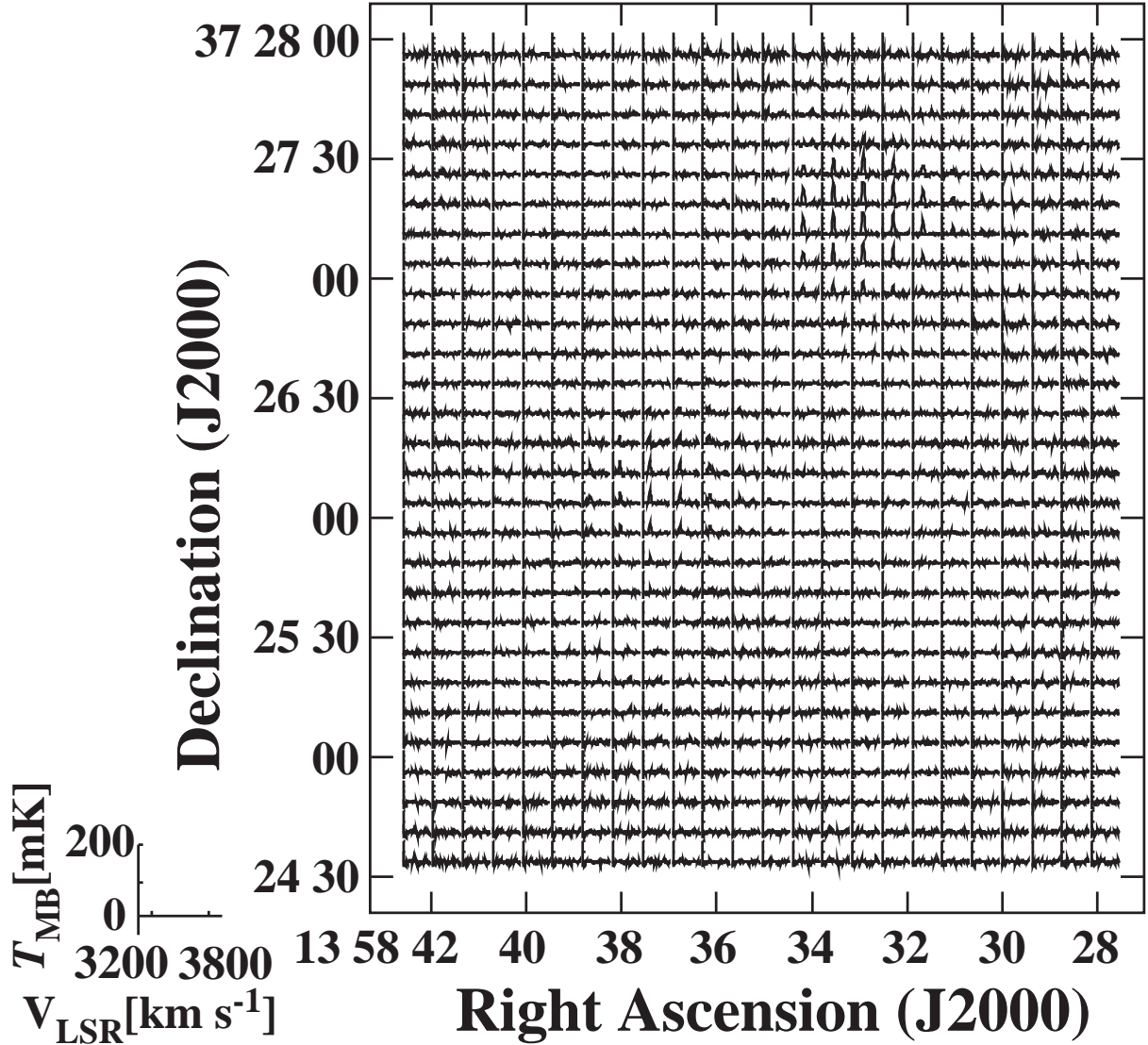}
		\caption{Spectrum map of CO(1-0) emission of Arp~84. The horizontal axis and vertical axis for each spectra are velocity in local standard of rest and main beam temperature. Mapped region is \timeform{210"} $\times$ \timeform{210"} which corresponds 54 kpc $\times$ 54 kpc in linear scale.}
		\label{arp84prof}
	\end{center}
\end{figure*}

\begin{figure*}
	\begin{center}
		\FigureFile(160mm,135mm){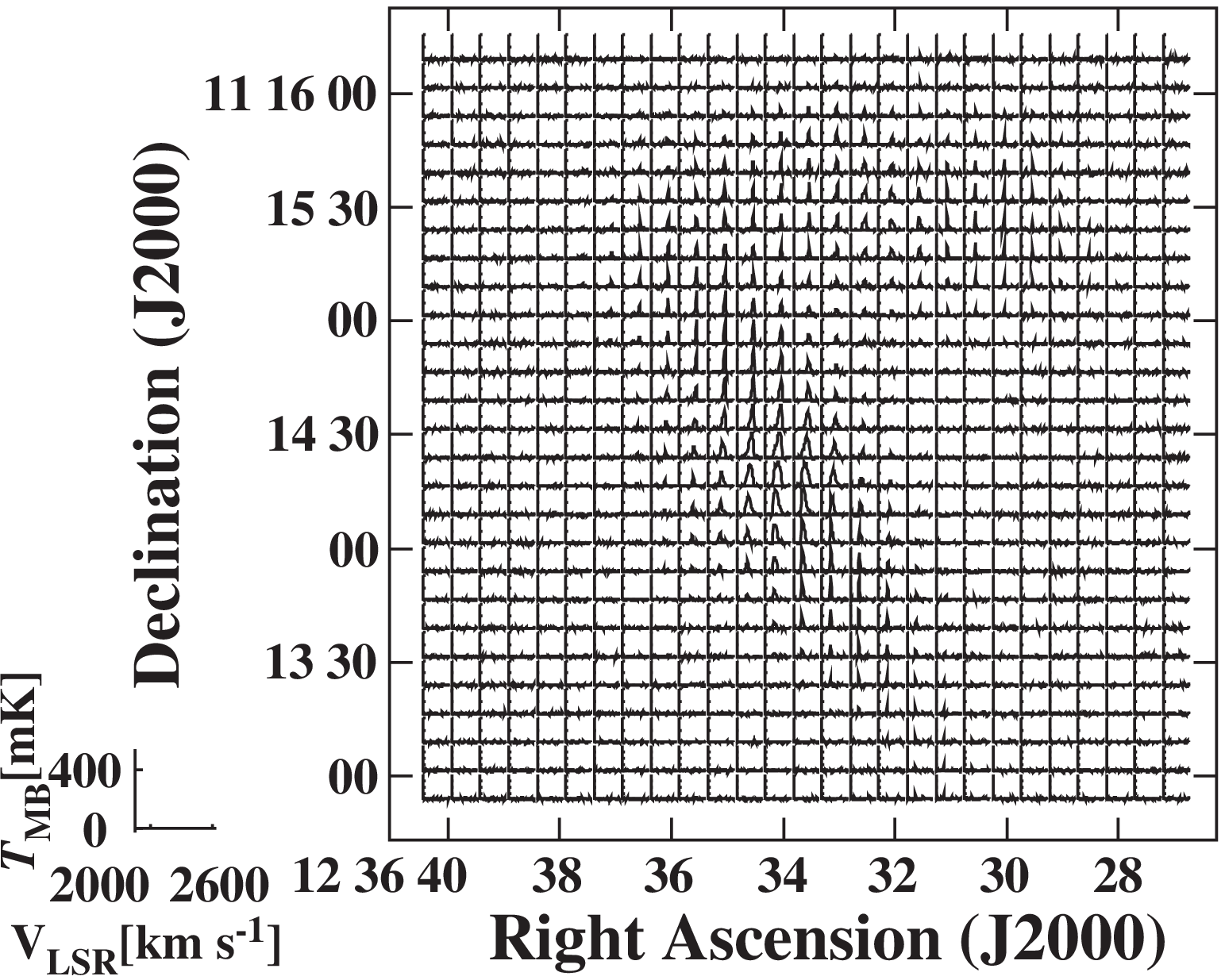}
		\caption{Spectrum map of CO(1-0) emission of VV~219. Mapped region is \timeform{210"} $\times$ \timeform{210"} which corresponds 16 kpc $\times$ 16 kpc in linear scale.}
		\label{vv219prof}
	\end{center}
\end{figure*}

\begin{figure*}
	\begin{center}
		\FigureFile(180mm,155mm){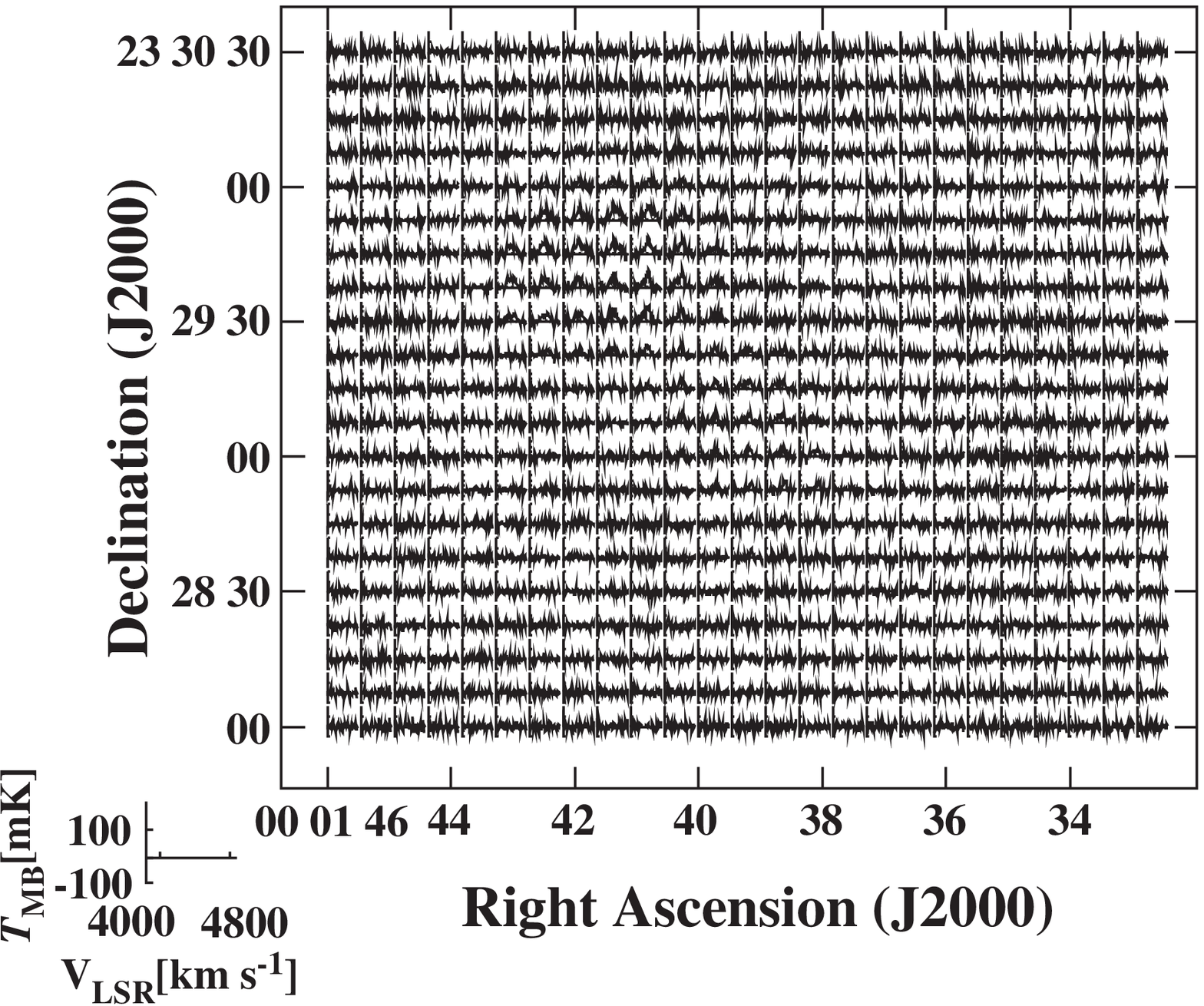}
		\caption{Spectrum map of CO(1-0) emission of VV~254. Mapped region is \timeform{210"} $\times$ \timeform{210"} which corresponds 63 kpc $\times$ 63 kpc in linear scale.}
		\label{taffy1prof}
	\end{center}
\end{figure*}

\begin{figure*}
	\begin{center}
		\FigureFile(160mm,130mm){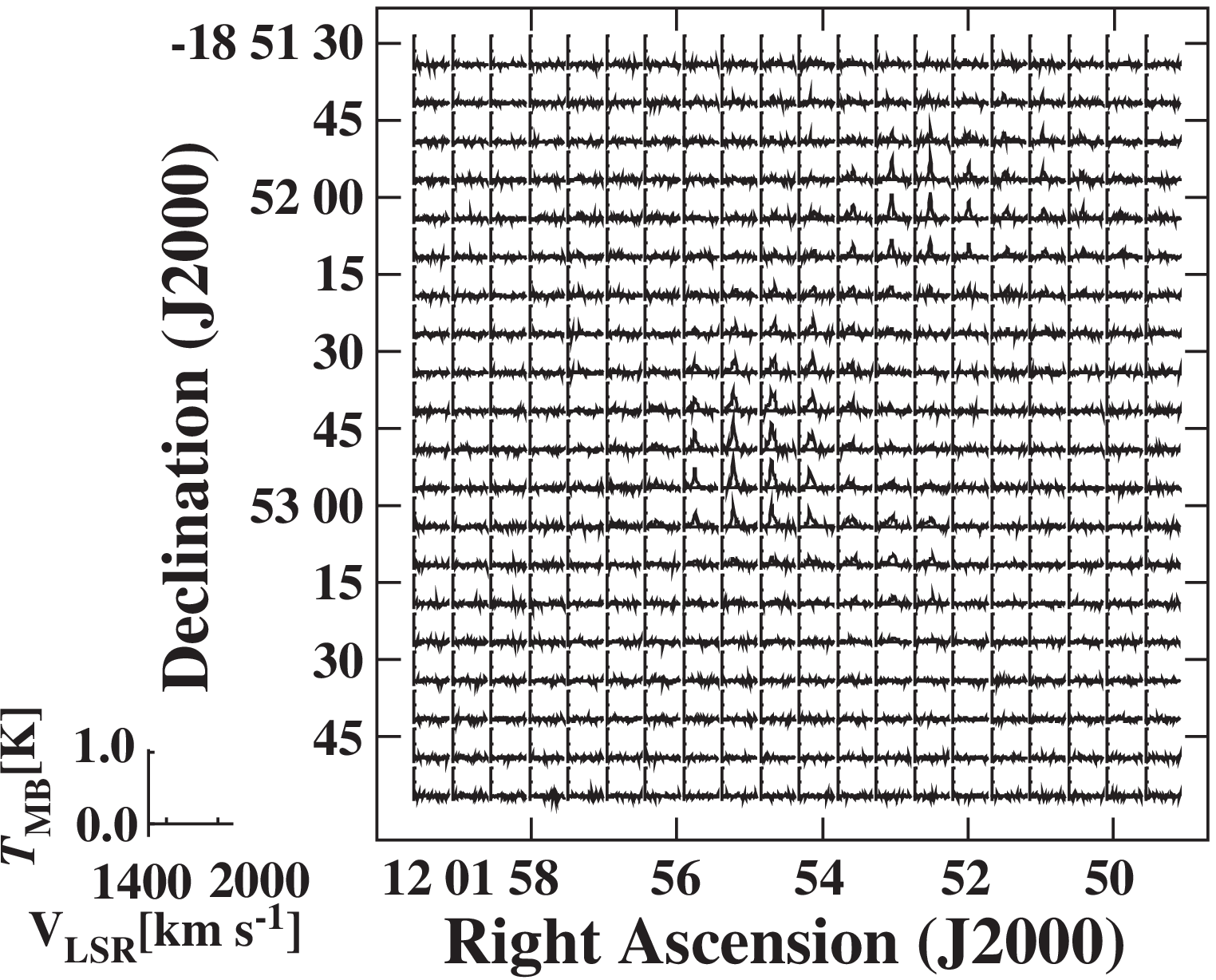}
		\caption{Spectrum map of CO(1-0) emission of the Antennae Galaxies. Mapped region is \timeform{150"} $\times$ \timeform{150"} which corresponds 15 kpc $\times$ 15 kpc in linear scale.}
		\label{antennaeprof}
	\end{center}
\end{figure*}

The spectra of $^{12}$CO emission of Arp~84, VV~219, VV~254 and the Antennae Galaxies obtained with the 45 m telescope are shown in Fig.\ref{arp84prof} --- \ref{antennaeprof}.

Achieved sensitivities were calculated by averaging the rms noise per channel of the spectra in emission free regions and are summarized in Table \ref{ObsRes}.
The uncertainties in the integrated intensity were estimated using the next equation;
\begin{equation}
	\Delta I = \sqrt{\Delta I_{n}^{2} + \Delta I_{b}^{2}}
\end{equation}
where
\begin{equation}
	\Delta I_{n} = T_{{\rm rms}} dv_{i}(dv_{c}/dv_{i})^{1/2}
\end{equation}

\begin{equation}
	\Delta I_{b} = T_{{\rm rms}} dv_{i}(dv_{c}/dv_{b})^{1/2}
\end{equation}
and $\Delta I_{n}$ the rms error in the spectrum, $\Delta I_{b}$ the error on the determination of the baseline.
In these equations, $T_{{\rm rms}}$ means the rms noise per channel and
$dv_{i}$ is velocity range over which the spectrum was integrated, $dv_{c}$ the smoothed channel width and $dv_{b}$ the velocity width over which the baseline was fitted.
Typically, $\Delta I_{n}$ is less than 10 \% of the total intensity and is mainly weather dependent.
The baseline uncertainty $\Delta I_{b}$ is usually 5-10 \%.
\begin{table}[tbp]
	\begin{center}
	\caption{Achieved $^{12}$CO(1-0) sensitivities}
	\begin{tabular}{ccc}
		\hline
		Source & $T_{{\rm rms}}$\footnotemark[$*$] & Observing time\footnotemark[$\dagger$]\\
		       & (mK)                & (hours)         \\
		\hline
		Arp~84 &  14  & 32\\
		VV~219 &  19 & 40\\
		VV~254 & 25 & 24\\
	    Antennae Galaxies &  37 & 10\\ 
		\hline
		\multicolumn{3}{@{}l@{}}{\hbox to 0pt{\parbox{85mm}{\footnotesize
			\footnotemark[$*$] rms noise per channel with 20 km s$^{-1}$ velocity resolution.
		\par\noindent
            \footnotemark[$\dagger$] the net observing time after subtracting bad data.\\
		}\hss}}
	\end{tabular}
	\label{ObsRes}
	\end{center}
\end{table}

\subsection{Comparisons with previous single-dish observations}
\label{COflux}
Past CO observations toward our sample were used to check our intensity calibration.
At first, we derived the flux from our $^{12}$CO integrated intensity map.
The formula to derive the flux density ($S_{\mathrm {CO}}$ in Jy km$^{-1}$) per pixel in maps from the brightness temperature ($T_{\mathrm {B}}$ in K) is as follows:
\begin{equation}
	S_{\mathrm{CO}} = \frac{2k}{\lambda^{2}}\ \Omega \ {\it T}_{\mathrm{B}}
\end{equation}
where {\it k} is the Boltzmann constant, $\lambda$ is the wavelength of the observed $^{12}$CO(1-0) line in rest frame (2.6 mm) and $\Omega$ is the solid angle of a pixel of our map\footnote{The solid angle occupied by a pixel (\timeform{7''.5} $\times$ \timeform{7''.5}) is 1.32 $\times$ 10$^{-1}$ steradian}. 
We adopted a conversion factor Jy K$^{-1}$ pixel$^{-1}$ is 0.5 and we use $T_{\mathrm {MB}}$ instead of $T_{\mathrm {B}}$.

The total CO flux of Arp~84 derived from our map including calibration uncertainties is 243 $\pm$ 38 Jy km s$^{-1}$ for NGC~5394 and 406 $\pm$ 63 Jy km s$^{-1}$ for NGC~5395.
These are consistent with the results of \citet{Zhu99} obtained with the NRAO 12-m telescope (201.3 $\pm$ 16.7 Jy km s$^{-1}$ for NGC~5394 and 547.8 $\pm$ 55.2 Jy km s$^{-1}$ for NGC~5395) 
within the error of their flux estimation (20-30 \%).
\citet{Zhu99} derived the total flux of NGC~5395 by assuming exponential radial distribution of molecular gas.
The real radial distribution appears steeper than the exponential model as seen in our map (Fig.\ref{arp84integ}(a)).
This may explain why the total flux of NGC~5395 by \citet{Zhu99} is larger than ours.
\citet{Kaufman99} derived the total flux of 177 Jy km s$^{-1}$ for NGC~5394 and 203 Jy km s$^{-1}$ for NGC~5395 using the 20-m telescope at Onsala Space Observatory.
Since they observed only 13 positions and they are not performed with Nyquist sampling, the flux is lower than our results.

The total flux of VV~219 system derived from our map is 714 $\pm$ 111 Jy km s$^{-1}$ for NGC~4567 and 1664 $\pm$ 258 Jy km s$^{-1}$ for NGC~4568.
Total CO flux of NGC~4567 and NGC~4568 reported by \citet{Young95} are 500 Jy km s$^{-1}$ and 1050 Jy km s$^{-1}$, respectively.
Since they observed only one position and their beam does not cover the entire region of the galaxies, lower flux is expected.
Although \citet{Chung09} mapped the entire region of the galaxy pair using the Five College Radio Astronomy Observatory (FCRAO) 14-m telescope, 
the total flux they derived is 50 \% lower than ours (1500 $\pm$ 170 Jy km s$^{-1}$). 
Since they found their flux is $\sim$25 \% smaller than that obtained by BIMA \citep{Helfer03} and attributed the discrepancy to the systematic error in calibration,
their values may be underestimated.

The total CO flux of VV~254 derived from our map is 1422 $\pm$ 220 Jy km s$^{-1}$.
This is consistent with the results obtained with the IRAM 30-m telescope by \citet{Braine03} (1246 Jy km s$^{-1}$) and 
those with the NRAO 12-m telescope by \citet{Zhu03} (1012.1 $\pm$ 171.6 Jy km s$^{-1}$), taking into account that our mapped region is larger than theirs.

Our total CO flux for the Antennae Galaxies is 3153 $\pm$ 426 Jy km s$^{-1}$, which is 10 \% lower than that derived by \citet{Gao01} (3595 Jy km s$^{-1}$) using the NRAO 12-m telescope, 
consistent with the FCRAO 14-m telescope \citep{Young95} (2844 Jy km s$^{-1}$), 
and 30 \% higher than that detected by \citet{Zhu03} (2370 Jy km s$^{-1}$).
The observed region of \citet{Zhu03} is smaller than our map although both observations were accomplished by the NRO 45-m telescope.
We confirmed that our spectra are consistent with those in \citet{Zhu03} within their observed region.
On the other hand, \citet{Gao01} observed a \timeform{4'} $\times$ \timeform{4'} region which is larger than ours (\timeform{2'.5} $\times$ \timeform{2'.5})
and weak emission is seen in the region where we did not observe.
Therefore, the main cause of the difference of the total flux in these observations is explained by the difference of the mapping areas. 

We derived the molecular gas mass of each galaxy pair from CO flux using the following formula \citep{KY88}:

\begin{equation}
	M_{\mathrm{H_{2}}} = 3.9 \times 10^{-17} \ X_{\mathrm{CO}} \ D^{2} \ S_{\mathrm{CO}} 
\end{equation}
where $M_{\mathrm{H_{2}}}$ is molecular gas mass in the unit of \MO, $X_{\mathrm{CO}}$ the N$_{\mathrm{H_{2}}}$/$I_{\mathrm{CO}}$ conversion factor and $D$ distance to the galaxy in Mpc.
We adopt a standard conversion factor $X_{\mathrm{CO}} = 1.8 \times 10^{20}$  cm$^{-2}$ (K km s$^{-1})^{-1}$ \citep{Dame01}.
On deriving molecular gas mass, we divide the galaxy pair into two constituent galaxies in the manner described below.
We fit two ellipses to reproduce the contour of 20 mag arcsec$^{-2}$ in {\it Ks}-band.
We regard the region inside of fitted ellipses as the extent of the galaxy.
For VV~219 and the Antennae Galaxies, since fitted ellipses overlap, we divide two galaxies with a line which crosses the points where two ellipses intersect.
The bridge region of VV~254 is defined as \citet{Braine03}.
Fitted ellipses are drawn in red in Fig.\ref{arp84integ}(d), Fig.\ref{VV219integ}(d), Fig.\ref{taffy1integ}(d) and Fig.\ref{antennaeinteg}(d).
Total CO flux and molecular gas mass of each galaxy are summarized in table \ref{flux}. 
All galaxies in our sample have molecular gas mass of more than 10$^{9}$ \MO.

It needs to notice, however, that $X_{\mathrm{CO}}$ varies between galaxies \citep{NK95} and is not constant even within a galaxy \citep{Casasola07}.
Observations of nearby galaxies revealed that $X_{\mathrm{CO}}$ is a function of metallicity \citep{Arimoto96, Wilson95, Leroy11}.
Most important fact of $X_{\mathrm{CO}}$ for our studies that (U)LIRGs most of which are interacting galaxies in the late stage of the interaction have lower $X_{\mathrm{CO}}$ than our Galaxy \citep{Solomon97, DS98}.
Although past studies attribute low conversion factor to warmer molecular gas temperature and larger CO line width and velocity dispersion \citep{MB88, DS98, Narayanan11}, the origin of low $X_{\mathrm{CO}}$ in interacting galaxies is still unknown.
$X_{\mathrm{CO}}$ factor in the Antennae Galaxies and VV 254 are also derived. 
The Antennae Galaxies have an one order lower $X_{\mathrm{CO}}$ than the standard conversion factor \citep{Zhu03}.
\citet{Zhu07} found that $X_{\mathrm{CO}}$ in VV 254 is about 4 times lower compared to the standard conversion factor.
This fact suggests that estimated molecular gas mass may be overestimated even for Arp 84 and VV 219 for which we do not have available $X_{\mathrm{CO}}$.

\begin{table*}
	\begin{center}
	\caption{Derived Properties of Molecular Gas}
		\begin{tabular}{ccccc}
			\hline
			Name      & Total Flux (Jy km s$^{-1}$) & $M_{\mathrm{H_{2}}}$ (10$^{9}$\MO) & $C_{20}$\footnotemark[$\dagger$]  & $C_{*}$\footnotemark[$\ddagger$] \\
			\hline
			Arp~84    & 649 $\pm$ 103 & 12.76 $\pm$ 2.00 & & \\
			NGC~5394  & 243 $\pm$ 38   & 4.77 $\pm$ 0.75 & 0.42 & 0.80 \\
			NGC~5395  & 406 $\pm$ 63  & 7.99 $\pm$ 1.25 & 0.28 & 0.43 \\
			\hline		
			VV~219    & 2378 $\pm$ 369 & 4.26 $\pm$ 0.66 & & \\
			NGC~4567  & 714 $\pm$ 111  & 1.28 $\pm$ 0.20 & 0.49 & 0.46 \\  
			NGC~4568  & 1664 $\pm$ 258 & 2.99 $\pm$ 0.46 & 0.48 & 0.62 \\
			\hline
			VV~254  & 1422 $\pm$ 220 & 38.4 $\pm$ 5.94 & & \\
			UGC~12914 & 380 $\pm$ 59  & 10.3 $\pm$ 1.60 & 0.36 & 0.68 \\
			UGC~12915 & 554 $\pm$ 86  & 14.9 $\pm$ 2.31 & 0.36 & 0.79 \\
			the Bridge & 488 $\pm$ 75 & 13.2 $\pm$ 2.03 & & \\
			\hline				
			Antennae Galaxies & 3153 $\pm$ 426 & 9.33 $\pm$ 1.26 & & \\
			NGC~4038  & 2080 $\pm$ 281 & 6.15 $\pm$ 0.83 & 0.45 & 0.48 \\
			NGC~4039  & 1073 $\pm$ 145 & 3.18 $\pm$ 0.43 & 0.46 & 0.50 \\
			\hline
			\multicolumn{5}{@{}l@{}}{\hbox to 0pt{\parbox{180mm}{\footnotesize
            \footnotemark[$\dagger$] Central concentration of molecular gas $C_{20}$ is defined by equation \ref{CentMol}. 
               \par\noindent
            \footnotemark[$\ddagger$] Central concentration of star $C_{*}$ is defined by equation \ref{CentStar}.
			}\hss}}
		\end{tabular}
	\label{flux}
	\end{center}
\end{table*}

\subsection{Distributions of molecular gas and comparison with other Wavelength}
\label{comparison}
We present distribution of molecular gas and velocity field of our sample galaxy pairs and
compare between the distribution of molecular gas and those of H\emissiontype{I}, {\it Ks}-band, H$\alpha$ or FUV and 8 $\mu$m or 24 $\mu$m.
H\emissiontype{I} data which are obtained with VLA by \citet{Iono05} except for the Antennae Galaxies which was taken by \citet{Hibbard01} 
and {\it Ks}-band 2MASS archival data \citep{Jarrett00} are used for the comparison.
H$\alpha$ data of VV~219 is acquired from \citet{Koopmann01} and for the Antennae Galaxies from \citet{Xu00}.
FUV data are taken from GALEX data archive, and 8 $\mu$m and 24 $\mu$m are from Spizter IRAC and MIPS data archives \citep{Smith07}.
The angular resolution of H\emissiontype{I} is summarized in Table \ref{h1res}.
Point spread functions of {\it Ks}, H$\alpha$, FUV, 8 $\mu$m and 24 $\mu$m are typically \timeform{2''}, \timeform{2''}, \timeform{4''}, \timeform{2''} and \timeform{5''}, respectively.
For each galaxy pair, integrated intensity map, velocity field,  H\emissiontype{I} integrated intensity map and {\it Ks}-band image, H$\alpha$, FUV, 8 $\mu$m and 24 $\mu$m are shown in Fig.\ref{arp84integ} --- \ref{antennaeinteg}.

\begin{table}[tbp]
	\begin{center}
	\caption{Angular Resolution of H\emissiontype{I} data used for comparison with our $^{12}$CO(1-0) observations}
		\begin{tabular}{cc}
				\hline
				\multicolumn{1}{c}{Galaxy Pair}       & {Angular Resolution (arcsec)}  \\
				\hline
				Arp~84		 & 17.8 $\times$ 16.3 \\
				VV~219		 & 19.9 $\times$ 14.0 \\
				VV~254		 & 18.0 $\times$ 18.0 \\
				Antennae Galaxies & 11.4 $\times$ 7.4 \\
				\hline
		\end{tabular}
	\label{h1res}
	\end{center}
\end{table}

\subsubsection{Arp~84}
\begin{figure*}
	\begin{center}
		\FigureFile(160mm,155mm){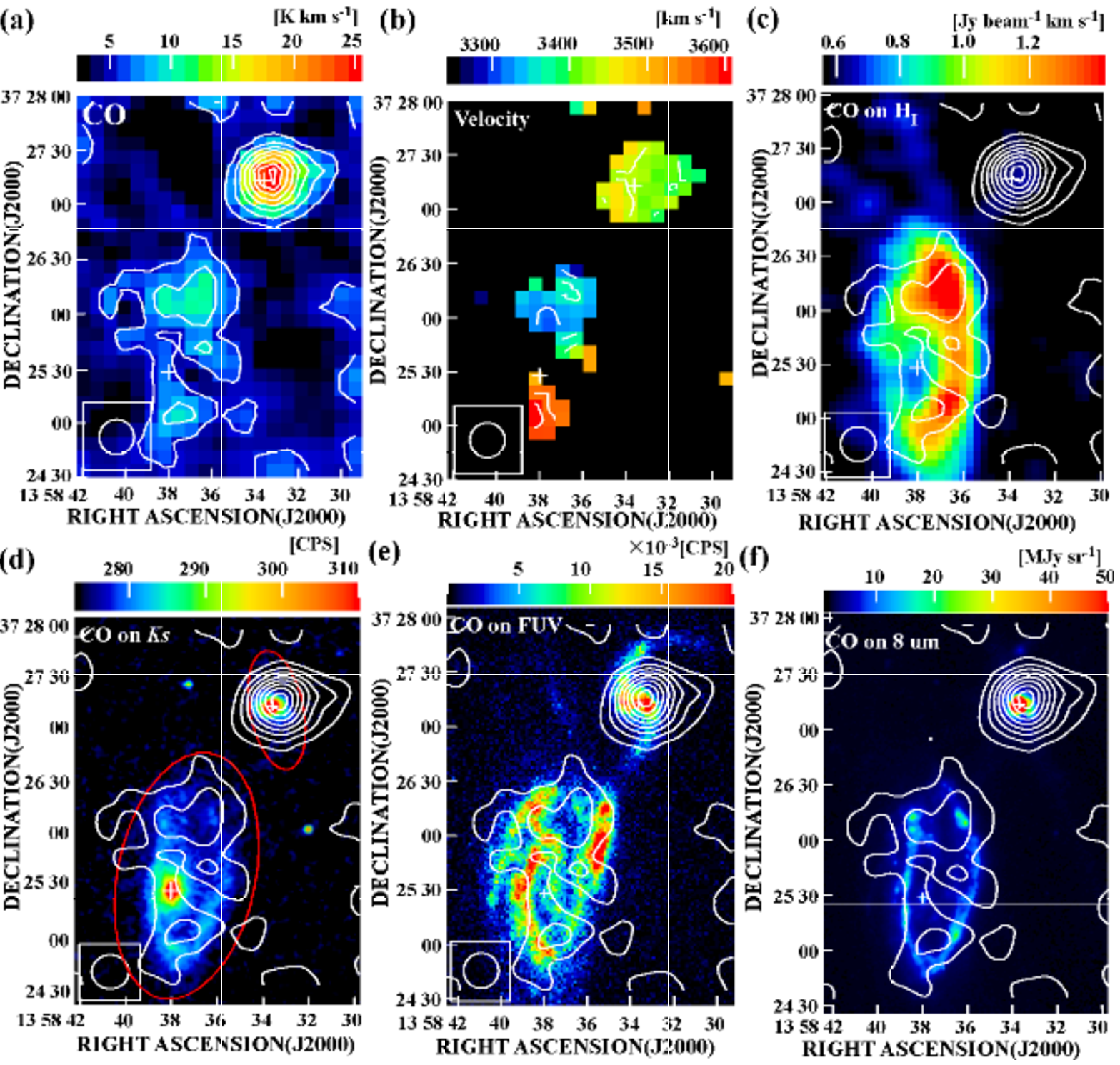}
		\caption{The crosses represent the galactic centre of NGC~5394 and NGC~5395.
			The contours are 2.81 $\times$ 2, 3, ... K km s$^{-1}$ for all images except for the image (d) where contours are  3160 + 40 $\times$ 1, 2, 3, ... km s$^{-1}$.
			(a) $^{12}$CO(1-0) integrated intensity map and contours. 
			(b) $^{12}$CO(1-0) velocity field.
			(c) $^{12}$CO(1-0) contours superposed on the H\emissiontype{I} image \citep{Iono05}. 
			(d) $^{12}$CO(1-0) contours superposed on the {\it Ks}-band image \citep{Jarrett00}.
			Red eclipses, result of the fit whose method is described in chapter \ref{COflux}.
			(e) $^{12}$CO(1-0) contours superposed on the GALEX--FUV image \citep{Bianchi03}.
			(f) $^{12}$CO(1-0) contours superposed on the MIPS 8 micron image \citep{Smith07}.}
		\label{arp84integ}
	\end{center}
\end{figure*}

Fig.\ref{arp84integ}(a) shows the CO integrated intensity map of Arp~84.
The CO map shows that NGC~5394 has a strong peak near the centre.
There is an offset between the position of the peaks of CO and {\it Ks} in NGC~5394. 
The CO peak is located about \timeform{5''} west of the {\it Ks} peak.
Although the offset is smaller than our beam size, the offset is also seen in the map obtained with an interferometer \citep{Iono05}.
The distribution of CO is slightly elongated from east to west in NGC~5394, while the {\it Ks} image shows the tidal tail from north to south.
H\emissiontype{I} is very weak in NGC~5394, while a strong peak is seen in CO.
This suggests that most of the interstellar gas exist as molecules in NGC~5394.
On the other hand, Fig.\ref{arp84integ} shows that NGC~5395 has extended structure in all data.
{\it Ks} image shows a strong peak at the centre and a long tidal arm from south to north.
H\emissiontype{I} and CO is distributed along the tidal arm.
H\emissiontype{I} also has weak tidal arm at the northern part of NGC~5395 where no CO emission is seen (Declination $>$ \timeform{37D27'00''}). 
There are H\emissiontype{I} and CO peaks at the northern end of the tidal arm, while there is no peak of CO and H\emissiontype{I} at the centre of NGC~5395 unlike NGC~5394.
Both NGC~5394 and NGC~5395 show the discrepancy between CO/H\emissiontype{I} and {\it Ks}.
The discrepancy may imply the different behavior of gas and stars in the interaction.
While the velocity field of NGC~5394 indicates that the galaxy is almost face on, NGC~5395 shows rotating features and weakly disturbed at the northern end of the tidal arm. 

Fig.\ref{arp84integ}(e) and (f) show the distribution of FUV and 8 $\mu$m both of which trace SF activity.
In NGC~5394, the distribution of FUV shows that SF activity is very high at the centre of the galaxy.
Moreover, star-forming regions are also found along the tidal spiral arms seen in {\it Ks}.
There is a local peak of SF in the middle of northern tidal arm of NGC~5394.
On the other hand, NGC~5395 shows an extremely complex distribution of star-forming regions.
Although a peak of SF activity is located near the peak of CO, it does not coincide with the CO peak.
The distribution of star-forming regions extends toward NGC~5394 and is distributed all over NGC~5395.
The distribution of star-forming regions traced by FUV shows that the tidal arm of NGC~5395 is connected with the northern tidal tail which is not seen in the distribution of CO and {\it Ks}-band.
The southern tidal arm of NGC~5395 traced by FUV extends to NGC~5394 and has its strongest peak in this galaxy.

The distribution of 8 $\mu$m of NGC~5394 shows a peak at the centre and in tidal spiral arms as seen in the FUV map.
Weak 8 $\mu$m emission from tidal spirals of NGC~5394 compared to FUV implies that the dust extinction at the tidal spirals of NGC~5394 is small.
The fact that star-forming regions traced by both FUV and 8 $\mu$m are seen at the centre of NGC~5394 suggests strong SF occurs and is concentrated to the centre of NGC~5394.
The distribution of 8 $\mu$m of NGC~5395 resembles to {\it Ks}-band.
Each local peaks of 8 $\mu$m in NGC~5395 are located near the CO peaks.

Discrepancies of distribution between interstellar gas (i.e., CO and H\emissiontype{I}) and star-forming regions (i.e., FUV and 8 $\mu$m) are also seen in field spiral galaxies, such as M~100 \citep{SG97}, NGC~5194 \citep{Calzetti05}, IC~342 \citep{Hirota10} and NGC~3147 \citep{Casasola08}.
In field spiral galaxies, the offsets between ISM and star-forming regions may be explained with timescale variations and SFE in the result of a spiral density wave.
Molecular gas forms stars in the density wave and then newly formed stars are separated from their parent molecular cloud.
While molecular gas moves to other condensation along the density wave, the young stars emit FUV and can heat only nearby dust.
In this scenario, CO peaks are regarded as the future SF site and decouple from star-forming tracers.
The inconsistency of distribution between ISM and star-forming regions in NGC~5395 is more eminent than that in spiral galaxies which is previously reported.
This fact may be explained that the disturbance and accumulation of ISM due to the interaction in NGC~5395 
is much faster than SF timescale and/or SF timescale becomes longer than the field spiral galaxies.
However, latter possibility is not likely to happen, since previous observations found that SFE, which is an inverse of SF timescale, in interacting galaxies is higher or at least the same compared to that in field spiral galaxies \citep{Young89, Casasola04}.
In either case, the CO peaks found in the edge of the northern tidal tale in NGC~5395 would be the birth site of new stars.

\subsubsection{VV~219}
\begin{figure*}
	\begin{center}
		\FigureFile(160mm,120mm){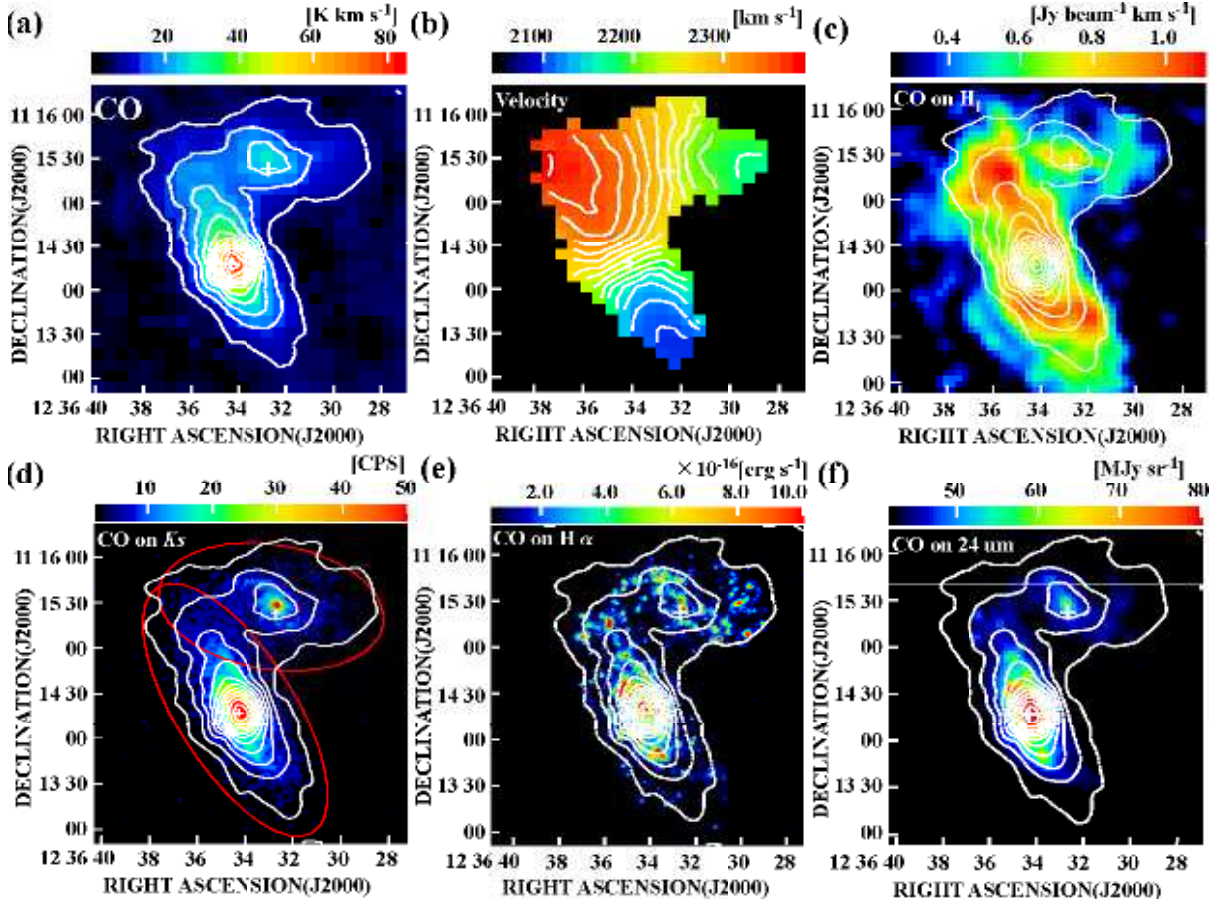}
		\caption{Same as Fig.\ref{arp84integ} but for VV~219 and (e) is $^{12}$CO(1-0) contours superposed on  the H$\alpha$ image and (f) is $^{12}$CO(1-0) contours superposed on  the MIPS 24 micron image. 
			The contours are 7.20 $\times$ 1, 2, 3, ... K km s$^{-1}$ for all images except for (b) where the contours are 2100 + 20 $\times$ 1, 2, 3, ... km s$^{-1}$.
			The crosses represent the galactic centre of NGC~4567 and NGC~4568.}
		\label{VV219integ}
	\end{center}
\end{figure*}

Fig.\ref{VV219integ}(a) shows the CO integrated intensity map of VV~219.
The figure shows that both galaxies have a central peak as often seen in nearby spiral galaxies (e.g., \cite{Kuno07}).
Although the distribution of molecular gas in these galaxies is not significantly disturbed by the interaction, 
the distribution of molecular gas in NGC~4568 shows asymmetry with respect to the minor axis of the galaxy.
That is, the extension of distribution of molecular gas on the northern side is larger connecting to NGC~4567 than the southern side.
The molecular gas smoothly connects from the northern edge of NGC~4568 to the eastern edge of NGC~4567.
The {\it Ks}-band image (Fig.\ref{VV219integ}(d)) shows spiral structure of both galaxies \citep{Jarrett03}.
Although the angular resolution of CO is not enough to resolve the spiral arms, we can see that the molecular gas tends to concentrate on the spiral arms.
NGC~4567 is known as a H\emissiontype{I} deficient galaxy \citep{KK04}.
In Fig.\ref{VV219integ}(c), most of the detected H\emissiontype{I} emission in NGC~4567 is distributed within the lowest contour of the CO map 
which corresponds to the surface density of molecular gas of 19 \MO pc$^{-2}$.
CO and H\emissiontype{I} extend toward the opposite direction in NGC~4568.
That is, while CO extends toward the northern side, H\emissiontype{I} extends toward the southern side.
The strongest H\emissiontype{I} peak is located on the elongated CO structure toward the north.

The velocity field of molecular gas is shown in Fig.\ref{VV219integ}(b).
It shows kinematics of molecular gas more clearly as compared with the previous studies whose field of view or angular resolution were limited (e.g., \cite{Iono05}).
The velocity fields of both galaxies are not significantly disturbed largely and overlap smoothly.
The radial velocity increases from west to east for NGC~4567 and from south to north for NGC~4568.
If we assume that NGC~4567/68 possess trailing arms, both galaxies rotate clockwise on the celestial sphere and are undergoing a retrograde-retrograde collision. 
Undisturbed but smooth connection at the distribution of molecular gas region in the velocity field confirms that NGC\,4567 and NGC\,4568 are not an apparent collision
but are really colliding.

As shown in Fig.\ref{VV219integ}(e) and (f), there are some similarities between H$\alpha$ and 24 $\mu$m distributions.
Massive star-forming regions are concentrated in the central region and spiral arms in both galaxies.
Although most of the star-forming regions are found along the spiral arms seen in {\it Ks} image (R $<$ \timeform{45''}), 
there are some large star-forming regions on the northern side of NGC~4568 which corresponds to the overlap region with NGC~4567, where strong H\emissiontype{I} peak and the CO elongation are seen.
These results imply that the interaction of NGC~4567 and NGC~4568 affected the SF in the overlap region.

\subsubsection{VV~254}
\begin{figure*}
	\begin{center}
		\FigureFile(160mm,120mm){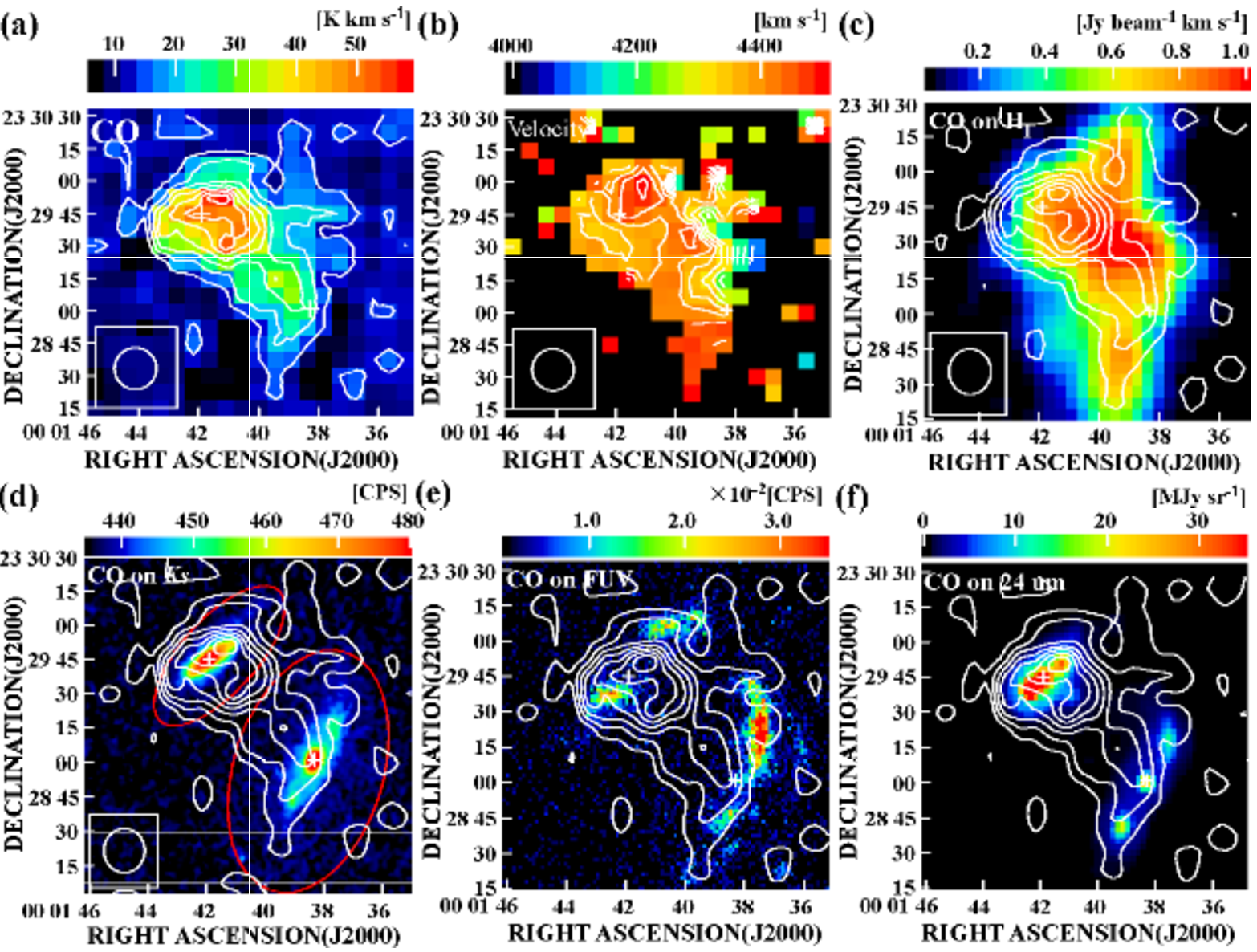}
		\caption{Same as Fig.\ref{arp84integ} but for VV~254 and  (e) is $^{12}$CO(1-0) contours superposed on  the GALEX--FUV image and (f) is $^{12}$CO(1-0) contours superposed on  the MIPS 24 micron image.
			The contours are 6.63 $\times$ 2, 3, 4, 5... K km s$^{-1}$ for all images except for (b) where the contours are 4100 + 20 $\times$ 1, 2, 3, ... km s$^{-1}$.
			The crosses represent the galactic centre for UGC~12914 (bottom right) and UGC~12915 (top left).}
		\label{taffy1integ}
	\end{center}
\end{figure*}

The CO integrated intensity map of VV~254 is shown in Fig.\ref{taffy1integ}(a).
Molecular gas in VV~254 is stretched out to the bridge region which is between the galaxies and its peaks are lopsided to the bridge region made by the head-on collision.
Although the apparent size of UGC~12915 is smaller than UGC~12914, a large amount of molecular gas belongs to UGC~12915.
The CO integrated intensity map also reveals that molecular gas has concentrated on the eastern side of UGC~12915. 

Fig.\ref{taffy1integ}(b) shows that kinematics of VV~254 is severely disturbed, especially in the bridge region.
Although both UGC~12914 and UGC~12915 are nearly edge-on, the velocity field does not show rotation.

Fig.\ref{taffy1integ}(c) shows that the distribution of atomic gas of VV~254 is much different from that of molecular gas.
It extends from north to south of VV~254 and most of the atomic gas is found in the middle of the bridge region.
Atomic gas is more rich in UGC~12914 than in UGC~12915 whereas molecular gas is more abundant in UGC~12915. 
Although the central region of two galaxies are not disturbed in {\it Ks}, the edges of the discs of the galaxies are warped toward the counterpart galaxy.

Fig.\ref{taffy1integ}(e) and (f) represents most star-forming regions in VV~254 are embedded in the discs of the galaxies and 
there is no star-forming region in the bridge region except for around the CO peak of UGC~12915.
In UGC~12914, the strongest SF activity traced by FUV shown in Fig.\ref{taffy1integ}(e) is found at the northwest disc where CO emission is weak.
Similarly, the most active star-forming region of UGC~12915 is located about \timeform{10''} from the centre and it does not correspond to the CO peak. 
Although both galaxies show weak SF activity in the centre of the galaxy in the FUV image,
24 $\mu$m image (Fig.\ref{taffy1integ}(f)) reveals that large star-forming regions are obscured in the centre of both galaxies.
UGC~12914 has three massive star-forming regions: the centre of the galaxy and both edge of the disc.
In the same way as FUV, the strongest star-forming region traced by 24 $\mu$m is seen at the western part of the disc of UGC~12915.

While molecular gas is plenty (a mass of more than 10$^{10}$ \MO) in the overlap region where large amount of atomic gas is found, there is no star-forming region there.
Since VV 254 is a head-on collision galaxy and is only 10$^{7}$ years after first encounter, almost all molecular gas could be gravitationally unbound.
The fact that the overlap region has higher $^{12}$CO(2-1)/ $^{12}$CO(1-0) and $^{12}$CO(3-2)/$^{12}$CO(1-0) ratio than UGC~12914 and UGC~12915 \citep{Zhu07} suggests that denser and warmer molecular gas which is more direct future SF tracer is embedded to the overlap region.
Therefore, VV 254 will produce stars at the overlap region in the future.

\subsubsection{The Antennae Galaxies}
\begin{figure*}
	\begin{center}
		\FigureFile(160mm,120mm){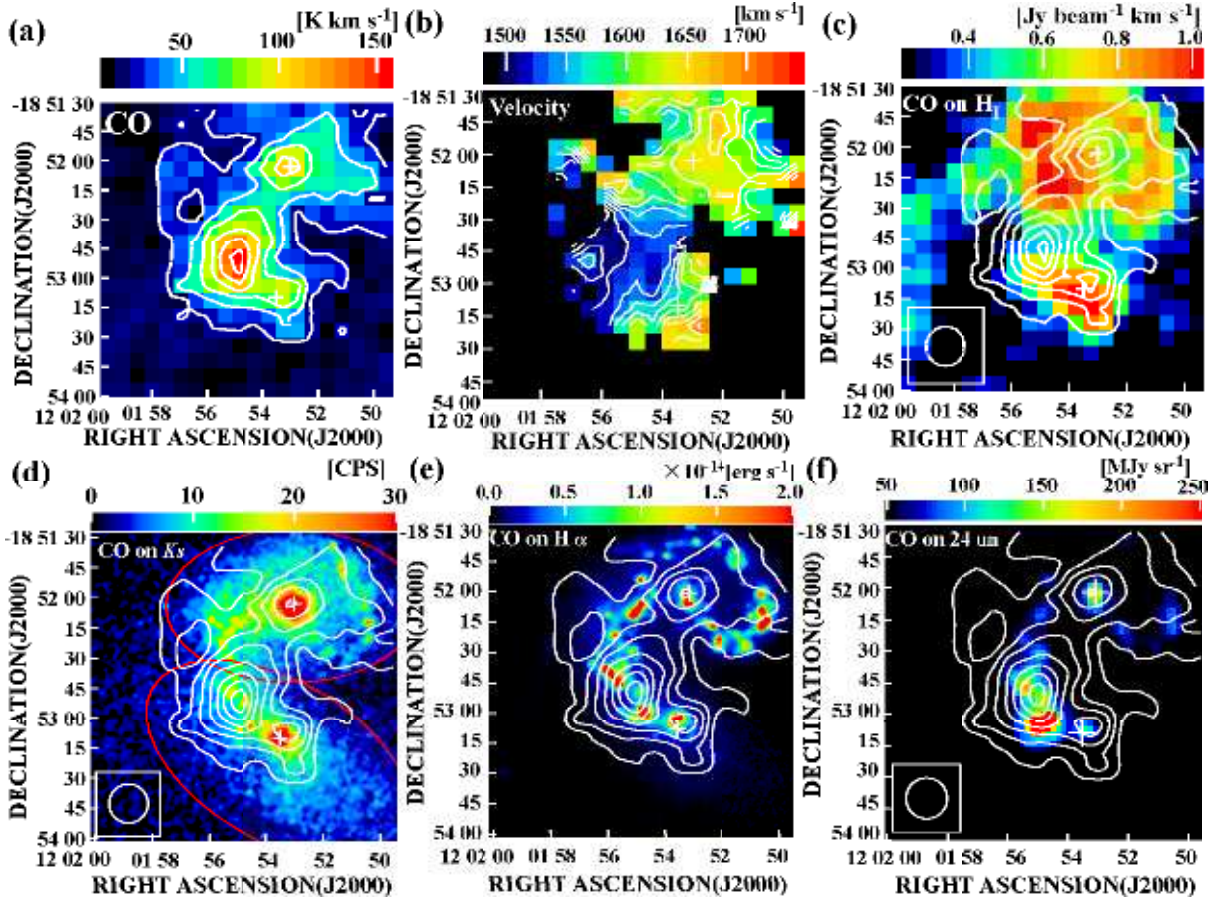}
		\caption{Same as Fig.\ref{arp84integ} but for the Antennae Galaxies and (e) is $^{12}$CO(1-0) contours superposed on  the H$\alpha$ image and (f) is $^{12}$CO(1-0) contours superposed on  the MIPS 24 micron image. 
				The contours are 23.7 $\times$ 1, 2, 3, ... K km s$^{-1}$ for all images except for (b) where the contours are 1500 + 20 $\times$ 1, 2, 3, ... km s$^{-1}$.
				The crosses represent the galactic centre for NGC~4038 (top) and NGC~4039 (bottom).}
		\label{antennaeinteg}
	\end{center}
\end{figure*}

Although a large amount of molecular gas exists in the centre of both galaxies, 
the largest concentration of molecular gas is seen in the overlap region (Fig.\ref{antennaeinteg}(a)). 
The mass of the concentration is more than 10$^{9}$ \MO.
We discovered molecular gas at the north east region around ($\alpha$, $\delta$) = (\timeform{12h1m57s}, \timeform{-18D52'10''}), which had not been found in the previous observations and belongs to the overlap region judging from Fig.\ref{antennaeprof}, 
although the amount of molecular gas in these structures is small.
On the other hand, atomic gas traces the tail structures and spiral arms of NGC~4038.
H\emissiontype{I} cavities are found at the centre of NGC~4038 and the overlap region where most molecular gas is found.
Additionally, the peaks of H\emissiontype{I} gas is located in at the north of the overlap region, the south of NGC~4039 and the tidal spiral arm of NGC~4038.
These regions do not correspond to the local peaks of CO emission.
Fig.\ref{antennaeinteg}(d) shows that old stars have smooth connections from NGC~4038 to NGC~4039.
Most molecular gas is distributed within the radius where brightness of {\it Ks}-band is 20 mag arcsec$^{-2}$ in the discs of these galaxies.
The velocity field shows complex kinematics of molecular gas (Fig.\ref{antennaeinteg}(b)).
While NGC~4039 shows nearly rotating features, the velocity field of NGC~4038 does not show clear kinematics.

H$\alpha$ image in Fig.\ref{antennaeinteg}(e) shows that there are many active star-forming regions whose luminosity is comparable to that in the central region of both galaxies.
These star-forming regions are distributed along the structures seen in {\it Ks}-band (Fig.\ref{antennaeinteg}(d)) and 
SF occurs not only in the region where CO emission is strong but also where CO is not strong like the northern distribution of molecular gas region.
Most star-forming regions traced by 24 $\mu$m is also found in H$\alpha$ image, although many star-forming regions are found in H$\alpha$.
Besides the distribution of star-forming regions, their relative luminosity differs between H$\alpha$ and 24 $\mu$m.
Although the distribution of star-forming regions traced by 24 $\mu$m coincides with CO distribution, 
the most active star-forming region is found in the southern distribution of molecular gas region which is located about \timeform{20''} (2 kpc in a linear scale) southeast from the CO peak.
Taking into account for a discrepancy of distributions between CO and H\emissiontype{I} like NGC~5395, the birth site of stars is expected to move to the northern part of the distribution of molecular gas region.

\section{Discussion}
\label{discussion}

In this section, we discuss central concentration of molecular gas in interacting galaxies by comparing them to isolated spiral galaxies to investigate the nature of the distribution of molecular gas under the interaction.

For the comparisons, the control sample of isolated spiral galaxies was obtained from Nobeyama CO atlas \citep{Kuno07}.
This atlas is the CO mapping survey of 40 nearby spiral galaxies using the NRO 45-m telescope.
Since the atlas was obtained with the same telescope as our data, the atlas data can be used for a direct comparison with our data.
To remove the environmental effect, we excluded the galaxies which belong to the Virgo Cluster and the Coma Cluster from the atlas sample.
We also removed galaxies whose H\emissiontype{I} data is not available.
The final control sample selected with these requirements comprises 20 galaxies summarized in Table~\ref{isolate}.

\begin{table*}[tbp]
	\begin{center}
		\caption{Control sample of isolated galaxies}
			\begin{tabular}{cccccccccc}
				\hline
				   Name    &   Morphology    &   Velocity\footnotemark[$*$]    &  Distance   & Inclination & $M_{\mathrm{H_{2}}}$\footnotemark[$\dagger$] & $R_{\mathrm{K20}}$\footnotemark[$\ddagger$] & $C_{20}$ & $C_{*}$ \\
				           &                 & (km s$^{-1}$) &    (Mpc)    &    (deg)    & (10$^{9}M_{\odot}$)  &     (arcsec)       &          &         \\
 				    (1)    &      (2)        &      (3)      &     (4)     &     (5)     &       (6)            &      (7)           &    (8)   &   (9)   \\
				\hline
				 NGC~253   &   SAB(s)c       &       227     &     2.5     &      75     & 1.6  & 630.2 & 0.78 & 0.77 \\
				 Maffei 2  &   SAB(rs)bc     &       -24     &     2.8     &      72     & 0.9  & 286.4 & 0.48 & 0.64 \\
				 IC 342    &   SAB(rs)cd     &        28     &     3.3     &      31     & 2.7  & 372.0 & 0.47 & 0.59 \\
				 UGC~2855  &   SABc          &      1207     &    10.4     &      63     & 20.3 & 114.3 & 0.53 & 0.73 \\
				 NGC~2903  &   SAB(rs)bc     &       549     &     6.3     &      67     & 1.6  & 163.0 & 0.54 & 0.62 \\
				 NGC~3184  &   SAB(rs)cd     &       594     &     8.7     &      21     & 1.1  & 114.6 & 0.38 & 0.57 \\
				 NGC~3351  &   SB(r)b        &       778     &    10.1     &      40     & 1.1  & 116.3 & 0.45 & 0.85 \\
				 NGC~3521  &   SAB(rs)bc     &       792     &     7.2     &      63     & 2.9  & 164.4 & 0.63 & 0.86 \\
				 NGC~3627  &   SAB(s)b       &       715     &    11.1     &      52     & 8.8  & 185.0 & 0.67 & 0.75 \\
				 NGC~3631  &   SA(s)c        &      1164     &    21.6     &      17     & 3.6  & 80.9  & 0.52 & 0.76 \\
				 NGC~4051  &   SAB(rs)bc     &       725     &    17.0     &      49     & 2.7  & 102.6 & 0.44 & 0.75 \\
				 NGC~4102  &   SAB(s)b       &       853     &    17.0     &      56     & 2.2  & 68.6  & 0.72 & 0.80 \\
				 NGC~4736  &   (R)SA(r)ab    &       317     &     4.3     &      40     & 0.4  & 172.3 & 0.91 & 0.87 \\
				 NGC~5055  &   SA(rs)bc      &       503     &     7.2     &      61     & 2.9  & 204.2 & 0.48 & 0.75 \\
				 NGC~5236  &   SAB(s)c       &       514     &     4.5     &      24     & 2.0  & 312.4 & 0.87 & 0.70 \\
				 NGC~5248  &   SAB(rs)cd     &      1165     &    22.7     &      40     & 8.8  & 111.8 & 0.68 & 0.70 \\
				 NGC~5457  &   SAB(rs)cd     &       255     &     7.2     &      18     & 3.2  & 236.3 & 0.44 & 0.80 \\
				 NGC~6217  &   (R)SA(rs)bc   &      1355     &    23.9     &      34     & 1.8  & 73.2  & 0.69 & 0.57 \\
				 NGC~6946  &   SAB(rs)cd     &        60     &     5.5     &      40     & 4.0  & 252.5 & 0.57 & 0.46 \\
				 NGC~6951  &   SAB(rs)bc     &      1425     &    24.1     &      30     & 5.9  & 115.6 & 0.59 & 0.62 \\
				\hline
				\multicolumn{9}{@{}l@{}}{\hbox to 0pt{\parbox{180mm}{\footnotesize
                  \footnotemark[$*$] Velocity in local standard of rest \citep{Kuno07}.
   	                  \par\noindent
                  \footnotemark[$\dagger$] Molecular gas mass from \cite{Kuno07}.
                      \par\noindent
                  \footnotemark[$\ddagger$] The radius at 20 mag arcsec$^{-2}$ in the {\it Ks}-band \citep{Jarrett03}.
                }\hss}}
			\end{tabular}
	\label{isolate}
	\end{center}
\end{table*}

One of the scenarios which explains the explosive SF activity found in interacting galaxies in the late stage of the interaction is gas infall toward the galactic centre as shown by \citet{BH96}.
The gas infall leads to an enhancement of the density of molecular gas in the central region of the galaxies and active SF will be triggered using the dense molecular gas.
In fact, high molecular gas concentration has been found in late stage interacting galaxies such as Arp~220 and NGC~520 whose SF is extremely active \citep{Scoville86,Sanders88}.

On the other hand, for early and mid stage interacting galaxies, 
\citet{Iono05} claimed that molecular gas is not concentrated to the nucleus of progenitors as expected in numerical simulations.
Their observations, however, were carried out with the interferometer and mapped only central regions of the sample galaxies.
The missing flux problem of the interferometer disables measurements of  the degree of concentration of molecular gas.
In addition, because of the limited observing area, it is difficult to know the actual distribution of molecular gas in the early and the mid stage of the interaction in which molecular gas is distributed over a relatively wide area in the galactic discs.
Hence, mapping observations of the whole region of the galaxy pair with a single-dish telescope are important to quantify concentration of molecular gas.

In this paper, molecular gas concentration is defined by 
\begin{equation}
	C_{20}=\frac{M_{\mathrm{H_{2}}}(R_{K20}/2)}{M_{\mathrm{H_{2}}}(R_{K20})},
	\label{CentMol}
\end{equation}
where $M_{\mathrm{H_{2}}}(R)$ is molecular gas mass within a radius $R$ and $R_{K20}$ is the radius at 20 mag arcsec$^{-2}$ in {\it Ks}-band.
We took $R_{K20}$ value from 2MASS Large Galaxy Atlas \citep{Jarrett03}.
For the galaxies which are not included 2MASS Large Galaxy Atlas, we derived $R_{K20}$ from the Two Micron All Sky Survey (2MASS) {\it Ks}-band archival data for isolated spiral galaxies 
\citep{Jarrett00}\footnote{This publication makes use of data products from the Two Micron All Sky Survey, 
which is a joint project of the University of Massachusetts and the Infrared Processing and Analysis Center/California Institute of Technology, 
funded by the National Aeronautics and Space Administration and the National Science Foundation.}.
$R_{K20}$ of interacting galaxies are taken from the major axis of the ellipse which is defined in \S\ref{COflux}.
For VV~219 and the Antennae Galaxies, since 20 mag arcsec$^{-2}$ contour connects between two constituent galaxies,
we exclude the regions outside the red line which divide two galaxies to calculate molecular gas concentration.
The radius at 20 mag arcsec$^{-2}$ in {\it Ks}-band for each interacting galaxy is shown in Fig.\ref{arp84integ}(d) --- Fig.\ref{antennaeinteg}(d) with white ellipses.
$C_{20}$ is listed in Table \ref{flux} and Table \ref{isolate}.

\begin{figure*}
   \begin{center}
      \FigureFile(160mm,95mm){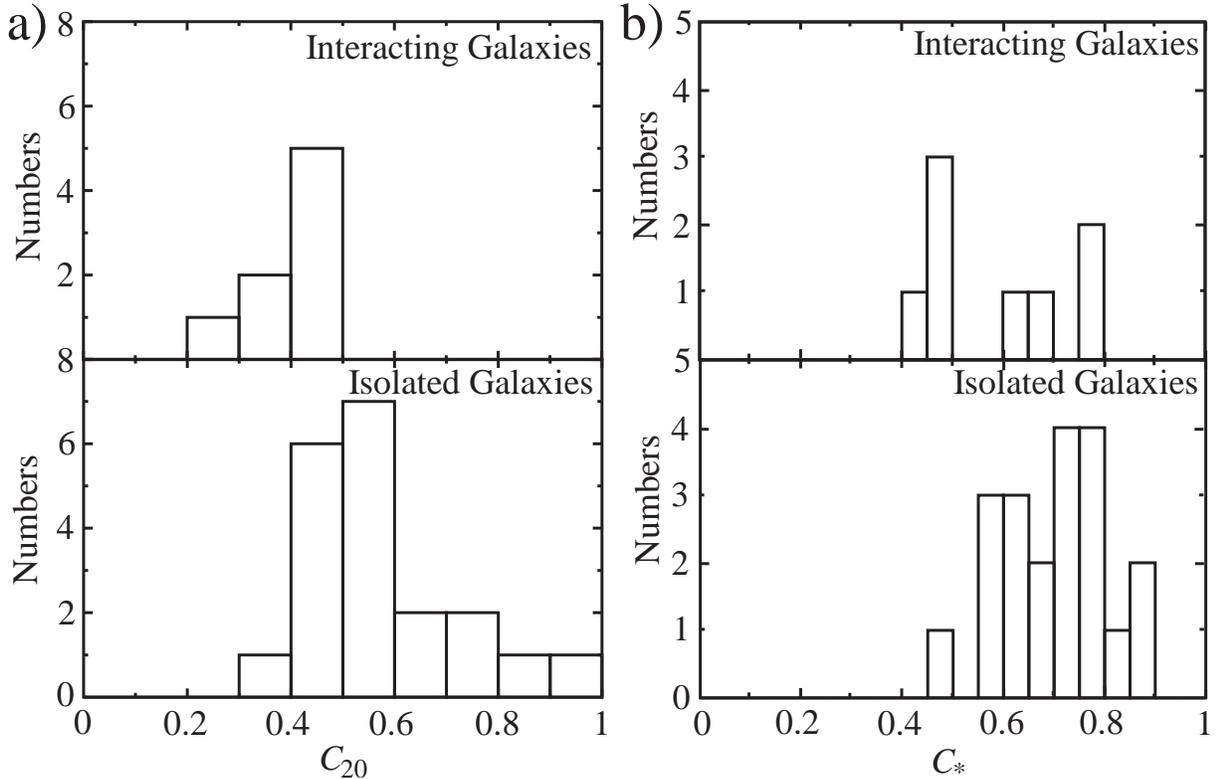}
   \end{center}
   \caption{(a) Histogram of $C_{20}$ for the interacting galaxies sample (upper panel) and the control sample (lower panel). (b) Histogram of $C_{*}$ for the interacting galaxies sample (upper panel) and the control sample (lower panel).}
   \label{C20hist}
\end{figure*}

Fig.\ref{C20hist}(a) is the histogram of $C_{20}$ for the interacting galaxies and the control galaxies sample.
The median of $C_{20}$ of interacting galaxies is 0.44 and the maximum is 0.49, while the median of the control sample is 0.56 and the maximum is 0.91.
We performed the Kolmogorov-Smirnov test between two samples and 
found that the probability that both samples are extracted from the same parent sample is 0.02.
As mentioned in \S\ref{comparison}, $C_{20}$ of NGC~5394 may be underestimated because CO extension is relatively small as compared to other galaxies.
Therefore we performed the Kolmogorov-Smirnov test assuming the $C_{20}$ value of 0.9 for NGC~5394 which is comparable with the maximum value of the control sample.
Even in this case, the possibility is 0.08.
This result implies that molecular gas is not accumulated toward the centre of interacting galaxies directly by the interaction and 
that there is a phase in which distribution of molecular gas is significantly disturbed.

For comparison, we also derived central concentration of stars, $C_{*}$, using a {\it Ks}-band image which traces stellar mass.
In the same manner as $C_{20}$, $C_{*}$ is defined as
\begin{equation}
	C_{*}=\frac{L_{\mathrm{*}}(R_{K20}/2)}{L_{\mathrm{*}}(R_{K20})},
	\label{CentStar}
\end{equation}
where $L_{\mathrm{*}}$ is {\it Ks}-band luminosity.
$C_{*}$ is listed in Table \ref{flux} and Table \ref{isolate}.
The median of $C_{*}$ of interacting galaxies is 0.56 and the maximum is 0.79, while the median of the control sample is 0.74 and the maximum is 0.87.
The result of the Kolmogorov-Smirnov test of 0.20 shows these two samples are from same parent,
 although the control sample has higher value of $C_{*}$.

These comparisons of $C_{20}$ and $C_{*}$ imply that both molecular gas and stars are disturbed by the interaction and that 
the interaction has a stronger effects on molecular gas than stars in the early stage of the interaction.
Our observational finding that distributions of molecular gas in the early stage of the interaction are larger than isolated spiral galaxies is
consistent with recent simulations which treat feedback from SF and precise physics of interstellar medium \citep{Bournaud10,Saitoh09,Teyssier10},
while previous simulations which adopt isothermal interstellar medium \citep{BH96,MH96} do not show a significantly disturbed distribution of gas.
This fact indicates that direct comparisons can be done between observation and simulation about not merely of distribution of gas but SF associated with the change of physical properties which are discussed in upcoming papers.

\section{Summary}
We made CO mapping observations of interacting galaxies in the early and the mid stages of the interaction with the 45-m telescope at NRO.
Our high quality CO maps reveal that distribution of molecular gas differs from atomic gas, old stars and SF tracers suggesting the strong influence of the interaction even in the early stage of the interaction.
Some interacting galaxies show that molecular gas is lopsided to the area between two constituent galaxies.
Furthermore, we compared central concentration of molecular gas between interacting galaxies and isolated galaxies and found that the central concentration of molecular gas in interacting galaxies is lower than that in isolated galaxies.
Old stars also show interacting galaxies have lower central concentration than isolated galaxies but the difference is not so clear as compared with that of molecular gas.
The results suggest that even in the early stage of the interaction the distribution of molecular gas is strongly disturbed by the interaction and there is a phase when the central concentration of molecular gas decreases before gas infall toward the galactic centre.\\

We would like to thank  the all the staff of the NRO for observational support.
This research has made use of the NASA/IPAC Extragalactic Database (NED) which is operated by the Jet Propulsion Laboratory,
California Institute of Technology, under contract with the National Aeronautics and Space Administration.
YT is supported by JSPS Grant-in-Aid for Research Activity Start-up (no. 23840007).

\end{document}